\documentclass[pre,twocolumn,preprintnumbers]{revtex4-1}

\usepackage{amsmath}
\usepackage{amssymb}
\usepackage{graphicx}

\newcommand{\Label}[1] {\label{#1}}
\newcommand{\reff}[1]{(\ref{#1})}

\newcommand{\Tensor}[1]{\mathbb{#1}}

\begin{document}

\title{Hydrodynamic modes in a confined granular fluid}

\author{Ricardo Brito$^1$, 
        Dino Risso$^2$, and 
        Rodrigo Soto$^{3}$ }
\affiliation{$^1$ Departamento de F\'{\i}sica Aplicada I (Termolog\'{\i}a), Universidad Complutense de Madrid, Spain\\
        $^2$ Departamento de F\'{\i}sica, Universidad
        del  B\'{\i}o-B\'{\i}o, Concepci\'on,  Chile\\
        $3$Departamento de F\'{\i}sica, FCFM, Universidad de Chile,
        Santiago, Chile}

\pacs {45.70.-n,  45.70.Mg }

\begin{abstract} 
Confined granular fluids, placed in a shallow box that is vibrated vertically, can achieve homogeneous stationary states thanks to energy injection mechanisms that take place throughout the system. These states can be stable even at high densities and inelasticities allowing for a detailed analysis of the hydrodynamic modes that govern the dynamics of granular fluids. 
Analyzing the decay of the time correlation functions it 
is shown that there is a crossover between a quasielastic regime in which energy evolves as a slow mode, to a inelastic regime, with 
energy slaved to the other conserved fields. The two regimes have well differentiated transport properties and, in the inelastic regime, the dynamics can be described by a reduced hydrodynamics with modified longitudinal viscosity and sound speed.
The crossover between the two regimes takes place at a wavevector that is proportional 
to the inelasticity.
A two dimensional granular model, with collisions that mimic the energy transfers that take place in a confined system is studied by means of microscopic simulations. The results show excellent agreement with the theoretical framework and allows the validation of hydrodynamic-like models.
\end{abstract}
\maketitle

\section{Introduction}

Granular fluids have become a prototype of non-equilibrium matter. The need of permanent energy injection to counter-balance the energy dissipation in the grain interactions, place these systems under permanent non-equilibrium conditions. Energy is injected through boundaries or external fields and it is dissipated at the small scale of grain-grain collisions. This fact violates the detailed balance condition necessary to reach equilibrium. It is then, one of the objectives in the study of granular fluids, the construction of valid statistical mechanics tools under these non-equilibrium conditions~\cite{RMP96,GoldRapid,BrilliantovPoeschel}.

The usual approaches to the study of granular fluids are kinetic theory (with different levels of approximation) or hydrodynamic-like models for the relevant fields \cite{BrilliantovPoeschel}. The dissipative nature of collisions implies that granular media cannot be simultaneously in homogeneous and stationary states and, typically, spatio-temporal structures develop \cite{Brito1,Brito2}. The homogeneous cooling state (HCS), in which the energy is non-stationary, has been widely studied showing that it becomes unstable in the long-wavelength regime \cite{HCS,InestabilidadHCS1,InestabilidadHCS2,BEEPL}. It is the reference state for developing kinetic  and hydrodynamic models of granular media with small or vanishing driving \cite{IHS3D1,IHS3D2,IHS3D3}. 
Conversely, stationary states can only be obtained by permanent energy influx. When granular media are driven by boundaries, typically large inhomogeneities develop even in the stationary regimes (see for example, \cite{Grossman}). Local energy balance can be  obtained by compensating the energy dissipation with shear heating. In this case it is of  particular interest the uniform shear flow (USF), in which all fields are uniform except for the velocity that shows a linear profile \cite{USFSantos, QHS}. As in the HCS case, the USF serves as a reference state to develop kinetic and hydrodynamic models. An important outcome is that the  transport coefficients for the linear dynamics close to the HCS and USF states are different \cite{USFSantos, QHS}. In both cases and in other studied states, however, a generic feature appears. The evolution of the energy shows two well differentiated regimes, depending on the dissipation \cite{TwoRegimes}. At low dissipations the energy evolves in long time scales and can be treated as another hydrodynamic field  in equal foot as the conserved fields (density and momentum). At large dissipations, on the other hand, the energy evolves fast and it is slaved to the density and velocity field. An example of this slaving is found in avalanches, in which the granular temperature is proportional to the velocity gradient squared in the so-called Bagnold scaling \cite{Bagnold,Campbell}. 

The crossover between the previous regimes, the quasielastic and the inelastic ones, is difficult to observe and characterize qualitatively. Only under dilute conditions, the quasielastic regime is observable at finite inelasticities. At moderate densities the inelasticity must be extremely small otherwise the only visible regime is the inelastic one \cite{TwoRegimes,Zippelius}. Related to this is the fact that in  dense or moderately dense regimes, granular fluids develop large inhomogeneities, when the use of hydrodynamic equations (with transport laws linear in the field gradients) are of questionable validity \cite{GoldRapid}. Some other authors extended the hydrodynamic description by using  nonlinear constitutive relations \cite{Dufty1,Dufty2}. Homogeneity can be achieved in small systems, in which the unstable wavevectors are not accessible . Again, the limitation to large wave vectors renders hydrodynamics of limit validity. In summary, the regime crossover has not been tested under  dense inelastic conditions, issue that is studied here.

The quasi two dimensional (Q2D) geometry offers a possibility to study this crossover and the properties of the hydrodynamic modes near stationary and homogeneous regimes. In this geometry, grains are placed in a box with large horizontal dimensions, while the vertical dimension is small, typically less that two diameters in height. When the box is vertically vibrated, grains get energy through the collisions with the top and bottom walls and this energy is then transferred to the horizontal degrees of freedom via grain-grain collisions. As these collisions are also inelastic, the system can achieve stationary states with finite energies. The vertical scale is fast and evolves in the scale of a few vibration periods. The horizontal dynamics, on the other hand, evolves in larger times scales characterized by the density and momentum conservation.
In this geometry, it is known that in a wide range of parameters including dense inelastic conditions, the system remains homogeneous in the horizontal directions \cite{olafsen,prevost2004,Melby2005,clerc2008}. The key element that allows for the establishment of stationary homogeneous states is that, for the effective horizontal dynamics, there is a distributed energy injection source. In the case of the Q2D systems, in the absence of friction, this energy source is Galilean invariant and conserves momentum locally.

In this article we study the hydrodynamic modes in a granular fluid with a distributed energy injection mechanism similar to the one in the Q2D geometry. The analysis, although inspired in the Q2D geometry, is generic and valid for three dimensional systems if a distributed energy injection mechanism is devised. It will be shown that there is a crossover between the quasielastic and inelastic regimes and the properties of the modes will be studied in detail in both regimes. The analysis will be done studying the density-density correlation functions that are obtained from fluctuating hydrodynamics. The intermediate scattering function and the dynamic structure factors provide information of the relevant modes and their time dependence. Finally, we present  a discrete microscopic  model in which grains can gain or dissipate energy at collisions and the results obtained from molecular dynamics simulation are analyzed under the described framework.

The plan of the paper is the following. In Sect.~\ref{sec.Skw.granular}  we develop the framework for the analysis of the hydrodynamic modes using correlation functions for a granular fluid. In Sections \ref{sec.inelastic.regime} and \ref{sec.quasielastic} the inelastic and quasielastic regimes and analysed in detail. In the inelastic case, we derive the temperature slaving that gives rise to a reduced hydrodynamics. Section \ref{sec.crossover} analyzes the crossover between these regimes, showing that it takes place in a wavevector proportional to the inelasticity. A microscopic collisional model in two dimensions, that mimics the Q2D dynamics is presented in Sect. \ref{sect.modelodelta}. Simulations of this model and comparison with the theoretical framework are shown in Sect. \ref{sec.md}. Finally, conclusions are given in Sect. \ref{sec.conclusions}.

\section{Hydrodynamic modes for granular fluids} \Label{sec.Skw.granular}

The goal of this section is to follow a procedure equivalent to that of Landau and Placzeck for 
granular fluids, considering the particular issues of such systems, 
like modification of the hydrodynamic equations, 
and deriving the modifications in both the intermediate scattering function  $F(k,t)$ and the dynamic structure factors  $S(k,\omega)$. 
Time correlation functions of equilibrium fluctuations are  standard tools in the study on fluids, 
as they  contain equilibrium properties
(like, e.g. specific heats, or the speed of sound) as well as non equilibrium ones (transport coefficients). 
Onsager's regression hypothesis states that spontaneous fluctuations in equilibrium obey the same evolution 
equations that describe the macroscopic relaxation of an external perturbation, provided that the perturbation 
is weak. As, in the hydrodynamic limit of long wave lengths,  macroscopic relaxation proceeds via the Navier Stokes 
equations, the correlation function also evolves according to those equations. Then, the correlation functions can be used
to measure transport coefficients and other thermodynamical properties of the fluid \cite{hansenMcdonald,Boon}.  

Let us define a general space and time correlation function between the 
dynamic variables $A$ and $B$ as
\begin{equation}\Label{CAB}
C_{AB}({\bf r},t)= \langle\delta A({\bf r}+{\bf r}',t+t') \delta B ({\bf r}',t')\rangle ,
\end{equation} 
where $\delta A ({\bf r},t)= A({\bf r},t)-\langle A({\bf r},t)\rangle$
is the fluctuation of the variable $A$ with respect to its average value. In the definition above, we have assumed that the system is spatially homogeneous and invariant under time translation, so the system must reach a stationary state; otherwise the correlation function would depend on ${\bf r}$ and ${\bf r}'$ and also on $t$ and $t'$. 
Although the definition for $C_{AB}$ is general, we will restrict to the study of density autocorrelation 
function, where $A=B=\rho$, or the velocity correlation function, where $A=B={\bf u}$, that leads to a tensorial correlation 
function. Definitions of the observables in term of microscopic quantities are given in Appendix \ref{app.Skw}.

For practical purposes, it is convenient to take the Fourier transform in space of $C_ {AB}({\bf r},t)$ 
to obtain the so called intermediate scattering function, denoted by $F(k,t)$. Furthermore, the Fourier transform in time can be taken to get the dynamic structure factor, $S(k,\omega)$, whose properties for equilibrium fluids are 
explained in Appendix  \ref{app.Skw}.  At long or hydrodynamic wavelengths, larger than the mean free path or the size of the molecules, 
the so called Landau-Plazceck approximation allows calculation of $F(k,t)$ and $S(k,\omega)$. Such calculation is based in the fact that the
time dependent hydrodynamic fields can be described by the set of Navier Stokes equations at linear level. 
In contrast with the molecular fluids, $S(k,\omega)$   has not been widely used for inelastic or 
dissipative systems, but only recently \cite{Zippelius,puglisi1,puglisi2}. 

In order to construct such functions, we need the evolution equations for the system.
We  consider granular particles with energy is dissipated at every collision,  so we will not consider here systems with a Stokes friction, like those of \cite{puglisi1,puglisi2}. Then, in order to reach a stationary state, we have to supply energy into the system. There are many models
for energy injection \cite{Twan,AlainBarrat,SantosMontanero,Zippelius,puglisi1,puglisi2}, and we will introduce a collisional model for energy injection in Sect.~\ref{sect.modelodelta}, but for the time being we will develop the theory as much as we can without specifying its detailed form. We will  assume that the thermostat does not inject momentum, but only energy, and it is Galilean invariant. 
Under such assumption, the equations for the density field and momentum density are those of usual fluids: the continuity equation and the Navier Stokes' one. 

Despite the equation for the energy (or granular temperature $T$) 
does not derive from a microscopic conserved quantity,   one can write a balance equation for 
it. Taking as a starting point the conservation equation for the energy for elastic fluids, 
we must add a term that accounts for the dissipation and the energy injection. 
We will denote such term in the temperature equation  by $G(\rho,T)$, where we make explicit the dependence on the density and the temperature. It will also depend on microscopic coefficients like, for instance, the coefficient of normal restitution $\alpha$, and  also on parameters that characterize the energy injection. 

The second modification comes from the constitutive relation for the 
energy flux \cite{Sela,Rodrigo-mu,IHS3D1,BrilliantovPoeschel}. It  includes the usual 
heat conduction term, given by  Fourier's law and besides, there is a contribution
proportional to the gradient of the density, that has no counterpart in molecular fluids. 
Then the heat flux reads
\begin{equation}\Label{heatflux}
{\bf q}= - \kappa\nabla T - \mu \nabla \rho.
\end{equation}

With such considerations into account, we can write the nonlinear temperature equation as 
\begin{align}\Label{NLT}
\partial_t T({\bf r},t)= -{\bf u}\cdot\nabla  T -\frac{T}{\rho c_V} \left(\frac{\partial p}{\partial T}\right)_\rho \nabla \cdot {\bf u} \nonumber \\
-\frac{\Tensor{P}'}{\rho c_V}:\nabla {\bf u}  + \frac{1}{\rho c_V}\nabla(  \kappa\nabla T + \mu \nabla \rho) -G(\rho,T) .
\end{align}
Here $c_V$ is the specific heat at constant volume, $p$ is the hydrostatic pressure and $\Tensor{P}'$ is the traceless part of the stress tensor. At the hydrodynamic level, we assume that the stress tensor is Newtonian, characterized by shear and bulk viscosities $\eta$ and $\eta_V$. The arbitrary minus sign in front of $G$ has been included for later convenience. 

Let us note again that this  equation does not derive from a microscopical 
conserved quantity, reflecting that 
the term that describes the dissipation and 
the energy injection, $G$,  does not derive from a 
flux term, and therefore is not proportional to a gradient. 

When the system of granular particles evolves, the temperature may reach 
a stationary value, which is a balance between the dissipation 
and the energy injection. We assume that there exists a homogeneous stationary state. We can calculate the stationary 
temperature $T^{st}$ by integrating over the whole system Eq.~(\ref{NLT}), where all terms under a spatial derivative vanish, 
arriving at the expression
\begin{equation}\Label{Tst}
G(\rho, T^{st})=0 .
\end{equation}
This equation defines the stationary temperature in terms of the density, and other parameters included in $G$, like dissipation or the energy injection, that sets the functional form of $G$.  

When studying fluctuations about the stationary state, we use Onsager's regression hypothesis. We linearize the evolution
equations around the stationary state characterized by a constant density, a vanishing velocity and the temperature
 $T^{st}$. Then, we define fluctuations around such state as 
\begin{align}\Label{fluct}
\rho({\bf r},t)&=\rho +\delta \rho({\bf r},t),\\
 {\bf u}({\bf r},t)&=\delta{\bf u}({\bf r},t),\\
T({\bf r},t)&= T^{st} + \delta T({\bf r},t).
\end{align}
Linearization around such steady state follows the usual procedure as for molecular fluids. The new term, 
$G(\rho,T)$ linearizes to first order as
\begin{equation}\Label{Glin}
G(\rho,T)\simeq G_\rho \delta \rho + G_T\delta T,
\end{equation}
where $G_X$ denotes the derivative of $G$ respect to the variable $X$ evaluated at the average density and the stationary temperature. 
In an elastic fluid, the terms  $G_X$ are absent. They are present only in dissipative media and, in fact, they are proportional to the inelasticity of the medium. In what follows, before giving any explicit form of the energy injection mechanism, we will refer to them as the dissipation terms.

Then, the set of linear equations in the Fourier variable $\nabla \to ik$ reads
$\partial_ t\Psi  = -\mathsf{M}\Psi  $. The vector $\Psi  $ contains the Fourier transform of the 
fields, and $\mathsf{M}$ is the pseudo hydrodynamic matrix, with expressions
\begin{align}\Label{DeltaM}
\Psi &=
\left({\begin{array}{c} \delta \rho({\bf k},t)\\
			\delta u_{\parallel}({\bf k},t)\\
			\delta u_\perp({\bf k},t)\\
			\delta T({\bf k},t)  \end{array}}\right)  ,\\
\label{MatrixM}\mathsf{M}  &= 
\left( {\begin{array}{cccc}
0 &    i k \rho &     0  &        0 \\
\frac{ikp_\rho}{\rho} & k^2 \nu_l  & 0 &  \frac{ikp_T}{\rho}\\
0 & 0 & k^2 \nu  & 0  \\
G_\rho+ \frac{k^2 \mu}{c_V \rho} & \frac{ i k T^{st}p_T }{c_V \rho} & 0 & G_T+\frac{k^2 \kappa}{c_V \rho}
\end{array}}\right) ,
\end{align}
where $p_X$ denotes the derivative of $p$ respect to $X$ evaluated at the average density and the stationary temperature, $\nu=\eta/\rho$ is the kinematic viscosity and $\nu_l = (\eta+\eta_V)/\rho$ is the longitudinal kinematic viscosity. 
As usual, we have decomposed the velocity field ${\bf u}({\bf k},t)$ into its longitudinal,
$ u_{\parallel}({\bf k},t) = {\bf \widehat k}\cdot {\bf u}({\bf k},t)$,
and transversal, $u_\perp({\bf k},t)= {\bf u}({\bf k},t)- {\bf \widehat k}\,u_{\parallel}({\bf k},t)$, parts. The transversal part is in fact, a $D-1$ dimensional vector, and so it is the matrix that contains their components. 

The matrix $\mathsf{M}$ is the modified  hydrodynamic matrix for granular  fluids. It differs 
from the hydrodynamic matrix for molecular fluids  in two elements, 
related with the temperature: the $(T,\rho)$-element includes the new transport coefficient $\mu$ coming from
Eq.~(\ref{heatflux}), and the 
term $G_\rho$, while the $(T,T)$-element contains  the term $G_T$. Such terms modify drastically the spectrum of the matrix $\mathsf{M}$.
The matrix also has the modified equation of state and transport coefficients for a granular fluid, but these only modify quantitatively the matrix elements.

Obtention of the time dependence of the fields, required in Eq. (\ref{CAB}), 
involves the diagonalization of the matrix $\mathsf{M}$. As the matrix
is not Hermitian there are two sets of orthonormal eigenvectors, right and left ones, given by
\begin{equation}\Label{Eigen}
\mathsf{M}\psi_i=\lambda_i\psi_i; \quad \phi_i \mathsf{M} =\lambda_i\phi_i; \quad \phi_i\cdot\psi_j = \delta_{ij},
\end{equation}
with components labelled by the superindex $\beta$, that can take the values:
$\beta=(\rho,\parallel,\perp,T)$, in a self-explanatory notation. Then, the solution for
the deviations at time $t$, denoted by $\Psi  (t)$ are
\begin{equation}\Label{Deltat}
\Psi   (t)=\sum_i e^{-\lambda_i t}\psi_i c_i ,
\end{equation}
where the coefficients $c_i$ are the projections of the fluctuations at initial time over the left eigenvectors
\begin{align}\Label{ci}
c_i&= \phi_i\cdot \Psi  (t=0) \nonumber \\
&= \phi_i^\rho \delta \rho({\bf k}) + 
\phi_i^\parallel \delta u_\parallel({\bf k}) + 
\phi_i^\perp  \delta u_\perp({\bf k}) + 
\phi_i^T  \delta T({\bf k}),
\end{align}
and the fluctuations without explicit dependence on time are evaluated at $t=0$. As we see, all the time dependence
is contained in the exponential terms, of those the Fourier transform will be taken to get $S(k,\omega)$.

As we are mainly interested in the density-density correlation function (see, however, Sec.~\ref{sect.transverse} when we also study the
transversal correlation function), we need to multiply the density fluctuation at time $t$ with that at time zero, 
obtaining
\begin{align}\Label{FF}
F({\bf k},t)&= \frac{1}{\rho V} \langle \delta\rho({\bf k},t) \delta\rho(-{\bf k},0)\rangle\\
&=\sum_i e^{-\lambda_i t}\psi_i^\rho \sum_\beta \phi_i^\beta S_{\beta\rho}(k),
\end{align}
where $V$ is the volume.
Here $S_{\beta\rho}(k)$ is the static (equal time) structure factors between the field $\beta$ and density field $\rho$.
In equilibrium, such structure factors are diagonal, that is, only the density-density term, $S_{\rho\rho}$,
contributes to the sum in Eq.~(\ref{FF}) \cite{hansenMcdonald}. However, in non-equilibrium fluids, the structure factors are not
diagonal \cite{Josechu}. For symmetry reasons scalar and vectorial fields do not couple, implying that 
$S_{\parallel\rho}=S_{\perp\rho}=0$, while  $S_{\rho T}\neq 0$. Such static structure factors 
can be calculated for granular fluids by using, e.g. the technique developed in \cite{Twan} for a `random kick' driving. 

Before doing the full diagonalization, we note that $\mathsf{M}$ is positive definite for small wavevectors,  that is all hydrodynamic modes are stable, as long as $G_T p_\rho>G_\rho p_T$, otherwise one mode becomes unstable. This instability is of van der Waals type, related to the negative compressibility of the reduced dynamics at small wavevectors (see the end of Sec.~\ref{sec.slaving}) \cite{Argentina,Cartes}. In Sec.~\ref{sect.modelodelta} we will show that this condition is always fulfilled for the energy injection method we devise.

From the structure of the pseudo-hydrodynamic matrix, the transverse mode decouples from the rest obtaining directly  the associate eigenvalue. The other three modes couple and give contributions to the dynamic structure factor. Considering the parity and complex structure of $\mathsf{M}$ it is possible to deduce that the eigenvalues have the form
\begin{align}
\lambda_{\pm} &= \pm i \omega_B(k) + \widetilde{\Gamma}(k) ,\\
\lambda_T &=\widetilde{D}_T(k) ,\\
\lambda_\perp & = \nu k^2 ,
\end{align}
where $\lambda_{\pm}$ are  the eigenvalues associated with the sound modes and $\lambda_T$ to the heat mode. Using the standard notation of elastic fluids $\widetilde{D}_T$ (even function  in $k$) is the dissipation rate of the thermal modes, $\widetilde{\Gamma}$ (even in $k$)  is the dissipation rate of the sound mode, and $\omega_B$ (odd in $k$)  is the frequency of the sound modes. 

In the Landau-Plazceck theory of elastic fluids, $\widetilde{D}_T=D_T k^2$, $\widetilde{\Gamma}=\Gamma k^2$ and $\omega_B=c_s k$, where  $D_T$ is the thermal diffusivity,  $\Gamma$ is the sound damping constant, and $c_s$ is the adiabatic sound velocity. In next sections, the eigenvalues of the inelastic model are computed and two regimes are clearly differentiated:  the 
so called dissipative regime (Sect. \ref{sec.inelastic.regime}), usual in granular media, and  the quasielastic regime (Sect. \ref{sec.quasielastic}). The latter  
deals with the elastic limit, to make connection with the usual hydrodynamics, and to verify that 
the elastic limit is a singular limit. The crossover is studied in Sect. \ref{sec.crossover}. Finally, comparison with molecular dynamics simulations of a collisional model is done.

Starting from Eq.~(\ref{FF}) and considering  the temporal parity of $F(k,t)$ and the presence of the three hydrodynamic modes that couple to the density, the intermediate scattering function can be written as 
\begin{widetext}
\begin{equation}
F(k,t)= S(k) \left[\left(1-\gamma^{-1}\right)e^{-\widetilde{D}_T t} + \left(\gamma^{-1} \cos(\omega_B t) + \frac{\widetilde{\Gamma}+(\gamma-1)\widetilde{D}_T}{\gamma \omega_B}\sin(\omega_B t)\right) e^{-\widetilde{\Gamma} t}\right] \Label{fit.Fkt}
\end{equation}
and, therefore, the dynamic structure factor is
\begin{align*}
S(k,\omega) =& S(k) \left[ 
2\left(1-\gamma^{-1}\right)\frac{\widetilde{D}_T^2}{\omega^2+\widetilde{D}_T^2} +
\right. 
\gamma^{-1} \left(\frac{\widetilde{\Gamma}}{(\omega+\omega_B)^2+\widetilde{\Gamma}^2} + \frac{\widetilde{\Gamma}}{(\omega-\omega_B)^2+\widetilde{\Gamma}^2}  \right) +\\
& \left. \frac{\widetilde{\Gamma}+(\gamma-1)\widetilde{D}_T}{\gamma \omega_B} \left(\frac{\omega+\omega_B}{(\omega+\omega_B)^2+\widetilde{\Gamma}^2} - \frac{\omega-\omega_B}{(\omega-\omega_B)^2+\widetilde{\Gamma}^2}  \right) \right],
\end{align*}
\end{widetext}
where $S(k)$ is the static (density-density) structure factor. These expressions are formally equal to their counterparts in elastic fluids \cite{Boon}, containing the Rayleigh, Brillouin and asymmetric peaks.
However, the factor $\gamma$, that in equilibrium is the adiabatic constant or the ratio of specific heats, contains here the static structure factor $S_{T\rho}$. 
Geometrically, $\gamma$   is defined such that  the area enclosed by the thermal peak (Rayleigh) divided by the area under the sound  peaks (Brillouin) is $\gamma-1$.

\subsection{Energy scaling and dimensionless variables}
To simplify the analysis, we consider the case of inelastic hard particles of diameter $\sigma$ and mass $m$,  characterized by a velocity-independent restitution coefficient. 
Before giving details of the specific implementation of the energy injection mechanism, we assume   that it introduces a unique energy scale. Fields  and time can be rescaled according to this energy scale. Instead we use  the equivalent procedure of rescaling to the stationary temperature $T^{st}$.
Therefore, the fields, transport coefficients, and eigenvalues rescale by dimensional arguments as:
$\rho\to\hat{\rho}$, ${\bf u}\to \sqrt{T^{st}} \widehat{{\bf u}}$, $T\to T^{st} \widehat{T}$, $\nu\to\sqrt{T^{st}} \widehat{\nu}$,  $\nu_l\to\sqrt{T^{st}} \widehat{\nu_l}$, $\kappa\to\sqrt{T^{st}} \widehat{\kappa}$, $\mu\to (T^{st})^{3/2} \widehat{\mu}$, $p_T\to \widehat{p_T}$, $p_\rho\to T^{st} \widehat{p_\rho}$, $G_T\to \sqrt{T^{st}} \widehat{G_T}$, $G_\rho\to (T^{st})^{3/2} \widehat{G_\rho}$, and $\lambda\to \sqrt{T^{st}} \widehat{\lambda}$.  
Finally, dimensionless magnitudes are defined by further normalization with appropriate powers of $m$ and $\sigma$ for each quantity.
To simplify notation, the hat symbol will be suppressed in what follows.

\section{Inelastic regime} \Label{sec.inelastic.regime}
The first regime we consider is the inelastic regime, in which dissipation is large enough such that the energy evolves faster than the density and momentum at the relevant, small, wavevectors. The eigenvalues in this regime are obtained as a series in $k$ keeping finite the dissipation parameters $G_\rho$ and $G_T$. The results, up to $k^2$ (i.e. hydrodynamic) order, are

\begin{align} 
\lambda_{\pm} &= \pm i  c_d k + \left(\frac{\nu_l}{2} + \frac{G_\rho p_T}{2 G_T^2} + \frac{p_T^2}{2c_V G_T \rho^2} \right) k^2 , \Label{mode.sound.inelastic}\\
\lambda_T &= G_T + \left(\frac{\kappa}{c_V \rho} -  \frac{G_\rho p_T}{G_T^2}  - \frac{p_T^2}{c_V G_T \rho^2} \right) k^2 ,\Label{mode.thermal.inelastic}\\
\lambda_\perp & = \nu k^2.
\end{align}
The thermal mode is drastically modified compared to elastic fluids, as contains a zero order contribution because of lack of energy conservation. Therefore, it is not a slow mode anymore. 

The sound mode contains the dissipative sound speed,  $c_d=\sqrt{p_\rho - p_TG_\rho/G_T}$, instead of the adiabatic velocity like in elastic systems (see Appendix \ref{appendix.quasielastic}). 
Again, it is a consequence of the lack of energy conservation. 
The dissipative  sound velocity corresponds to the isothermal velocity $\sqrt{p_\rho}$, corrected by the instantaneous coupling between $T$ and $\rho$ given by the constraint (\ref{Tst}). 
The sound modes are stable as long as $p_\rho G_T > p_TG_\rho$ as was previously indicated in Section \ref{sec.Skw.granular}. The other modes are unconditionally stable for small wavevectors.
The viscous mode is not coupled to the dissipation terms and therefore is the same as in the elastic case. 

Note that the sound velocity, the sound damping constant and the $k^2$ term in the thermal mode do not approach the elastic values when the dissipation vanishes ($G_\rho\to0$ and $G_T\to0$).  The elastic limit is singular and a detailed scaling must be performed to match both regimes. This scaling is done in Section \ref{sec.quasielastic} and the numerical demonstration of the crossover between the regimes is done in Section \ref{sec.md}.

Finally, let us mention that the new transport coefficient $\mu$ does not appear in the eigenvalues  (\ref{mode.sound.inelastic})-(\ref{mode.thermal.inelastic}) of the matrix $\mathsf{M}$, up to order $k^2$. Therefore, one cannot measure such transport coefficient with the method developed in this paper, but has to devise different methods \cite{Rodrigo-mu}.

\subsection{Slaving of the temperature field} \Label{sec.slaving}
Being the temperature a fast field, it is possible to slave it to the density and velocity fields and obtain a simpler dynamics. The linearized temperature equation reads
\begin{equation}
\partial_t \delta T = -\left(G_\rho + \frac{k^2 \mu}{c_V \rho} \right)\delta\rho - \frac{i k p_T}{c_V \rho} \delta u_{\parallel} 
- \left(G_T + \frac{k^2 \kappa}{c_V \rho} \right) \delta T. \Label{eq.Tfast}
\end{equation}
At the time scale  of the slow modes the temperature has relaxed to the stationary solution of \reff{eq.Tfast} allowing to obtain an explicit expression for the temperature field. As $\delta T$ enters as a spatial derivative in the momentum equation (\ref{DeltaM}) only terms up to order $k$ must be retained to reproduce hydrodynamic modes, giving $\delta T = -\frac{G_\rho}{G_T} \delta\rho - \frac{i k p_T}{c_V \rho G_T} \delta u_{\parallel}$. However, when substituted in the momentum equation the resulting eigenvalues differ to the sound modes (\ref{mode.sound.inelastic}) in terms of order $k^2$. The reason is that this simple Markovian slaving does not consider all contributions to the same order in $k$. 
Formally, equation (\ref{eq.Tfast}) can be written as
\begin{equation}
\partial_t \delta T = - \gamma \delta T + S(t) ,
\end{equation}
with $S=-\left(G_\rho + k^2 \mu/(c_V \rho) \right)\delta\rho - \frac{i k p_T}{c_V \rho} \delta u_{\parallel}$ and $\gamma=\left(G_T + k^2 \kappa/(c_V \rho) \right)^{-1}$. Its solution is
\begin{align}
\delta T(t) &= \int_0^{\infty} dt'\, e^{-\gamma t'} S(t-t') \nonumber\\
&= \int_0^{\infty} dt'\, e^{-\gamma t'} \left [S(t) - \dot{S}(t) t' + \ldots\right]\nonumber\\
&= \frac{S(t)}{\gamma} - \frac{\dot{S}(t)}{\gamma^2} + \ldots ,
\end{align}
where the first term gives the Markovian slaving. To evaluate the second term, the hydrodynamic equations for $\delta\rho$ and $\delta u_\parallel$ are used, noting that only terms up to order $k$ must be retained. A third term, proportional to $\ddot{S}$ gives contributions of order $k^2$, not being relevant to the hydrodynamic model. The resulting slaving for $\delta T$ is 
\begin{equation}
\delta T =  -\frac{G_\rho}{G_T} \delta\rho - \frac{i k p_T}{c_V \rho G_T} \delta u_{\parallel} -\frac{i k \rho G_\rho}{G_T^2} \delta u_{\parallel}
\end{equation}
and the reduced hydrodynamic matrix for the slow variables $\Psi=\left({\begin{array}{c} \delta \rho({\bf k},t)\\
			\delta u_{\parallel}({\bf k},t) \end{array}}\right)$ is
\begin{align}
\mathsf{M}  &=  
\left( {\begin{array}{cc}
0 &    i k \rho \\
ik \left(\frac{p_\rho}{\rho} -\frac{G_\rho p_T}{G_T \rho}\right) & 
\left( \nu_l +\frac{G_\rho p_T}{G_T^2}+\frac{p_T^2}{cV G_T \rho^2} \right) k^2
\end{array}}\right) .
\end{align}
Using this reduced hydrodynamic model, the sound modes are correctly reobtained.
Note that in the reduced hydrodynamics, there is an effective longitudinal viscosity that is the bare longitudinal viscosity modified by a term that depends on the energy injection-dissipation mechanism. 

In this reduced dynamics, it is clear the effect of the stability condition $p_\rho G_T > p_TG_\rho$. It guarantees that the sound modes are stable. When it is not fulfilled, a spinodal decomposition develops and non-linear terms are necessary to describe the long term dynamics, resulting in the van der Waals normal form \cite{Argentina,Cartes}.

\section{Quasielastic limit} \Label{sec.quasielastic}
When dissipation is small it is expected that the temperature field becomes a slow field together with the density and momentum fields.  The dissipation terms $G_X$  must be compared with the terms proportional to $k^2$ in (\ref{mode.thermal.inelastic}). Therefore, the scaling that  capture this quasielastic regime must be done in the dissipation together with the wavevector. It is obtained doing $G_\rho= \epsilon^2 \widetilde{G}_\rho$, $G_T= \epsilon^2 \widetilde{G}_T$, and $k=\epsilon \widetilde{k}$ ($\epsilon$ is a formal small parameter) and computing the eigenvalues as series in $\epsilon$. Keeping terms up to order $\epsilon^2$ and going back to the original variables the eigenvalues are
\begin{align} 
\lambda_{\pm} &= \frac{p_T\left(G_Tp_T/c_V\rho^2+G_\rho\right)}{2c_s^2}  
\pm i  c_s k + \Gamma k^2 ,\\
\lambda_T &= \frac{\left(G_Tp_\rho-G_\rho p_T^2\right)}{c_s^2} 
 + D_T k^2 ,\\
\lambda_\perp & = \nu k^2 ,
\end{align}
where 
\begin{align}
c_s&=\sqrt{p_\rho+p_T^2/c_V\rho^2}\\
\Gamma&=\frac{\nu_l}{2}+\frac{\kappa p_T^2}{2c_V\rho(p_T^2+p_\rho \rho^2 c_V)}\\
D_T&=\frac{\kappa p_\rho\rho}{p_T^2+c_Vp_\rho\rho^2}
\end{align}
are the elastic adiabatic sound speed, sound damping constant and the thermal diffusivity, respectively. 

The $k$-independent term in the sound modes could lead to the erroneous impression that they are fast modes and do not correspond to conserved fields. This is not the case as it should be recalled that the quasielastic regime is obtained when the dissipation  scales as $k^2$ and therefore the $k$-independent terms vanish also when the wave vector goes to zero. A detailed analysis of the crossover of the two regimes and a geometric interpretation of the scaling is given in the next section.

Note that, as in the inelastic regime, the eigenvalues do not depend on the transport coefficient $\mu$, up to order $k^2$.

\section{Crossover wavevector} \Label{sec.crossover}
The two regimes described in the previous Sections (inelastic and quasielastic) lead to different eigenvalues of the hydrodynamic modes. We make a special remark on the sound velocity which is either  isothermal or adiabatic. For a given dissipation, there is a crossover wavevector $k_0$, such that when $k\ll k_0$ the dynamics is inelastic while if $k\gg k_0$ the dynamics is quasielastic  (see Fig. \ref{fig.regimes}).

To compute $k_0$, we consider the full expression of the eigenvalues associated with the sound mode and get the real part of the eigenvalues, $\widetilde{\Gamma}$. For a given small dissipation,  $k_0$ is the wavevector in which $\widetilde{\Gamma}$ has changed appreciably from the $k=0$ limit. It can be verified by doing a full diagonilization of the pseudo-hydrodynamic matrix that, in the limit of small dissipations, $\widetilde{\Gamma}\approx \Gamma_0-\Gamma_4 k^4+\ldots$. 
Therefore, the crossover wavevector is computed as
\begin{equation}
k_0^4 \equiv -\left(\frac{24 \widetilde{\Gamma}}{\frac{\partial^4 \widetilde{\Gamma}}{\partial k^4}}\right)_{k\to0} .\Label{k0.pred}
\end{equation}
In the limit of small dissipation (i.e. the dimensionless non-hydrodynamic terms  $G_T$ and $G_\rho$ are small) $k_0 \propto G_X$. The  crossover wavevector is proportional to the dissipation with a proportionality constant that depends on the equation of state, the heat capacity and the ratio $G_\rho/G_T$.

The regimes are presented schematically in Fig. \ref{fig.regimes}.
In the elastic limit, where $G_X\to 0$, only the quasielastic regime, in the appropriate limit, is a valid description. When the dissipation is finite, the relevant regime depends on $k$, obtaining the inelastic regime in the limit of large systems ($k$ going to zero). The figure also shows the singular character of the elastic limit:  it is not possible to obtain it for finite dissipations by taking $k\to0$. It can only be obtained if the inelasticity is reduced simultaneously as shown in the scaling of Section \ref{sec.quasielastic}.

\begin{figure}[htb]
\begin{center}\includegraphics[width=.8\columnwidth]{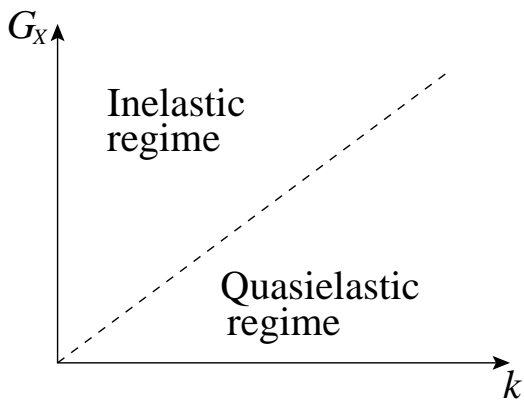}\end{center}
\caption{Schematic representation of the quasielastic and inelastic regimes, in terms of the wavevector $k$ and dissipation $G_X$. The crossover wavelength is $k_0\propto G_X$. At finite dissipation,  both regimes are present, depending on the wavevector. Only in the elastic case, the quasielastic regime is valid for all wavevectors.}
\Label{fig.regimes}
\end{figure}

\section{Collisional model} \Label{sect.modelodelta}
A particular geometry that has gained interest in the study of granular media is the quasi two-dimensional one (Q2D). Here, the box is large in the horizontal directions, while the vertical one is smaller than two particles' diameter, such that grains cannot be on top of another. If the box is vertically vibrated, with a maximum acceleration larger than gravity, grains gain vertical energy by collisions with the top and bottom walls and this energy is transferred to the horizontal directions through grain-grain collisions. Seeing from above the granular system is fluidized and can remain homogeneous under a large range of parameters. Varying the vibration amplitude and frequency the system develops a phase transition mediated by waves~\cite{clerc2008}
with a solid-like region coexisting with the fluid \cite{olafsen,vega}. Here we focus on the homogeneous state and with that purpose an effective two-dimensional model is proposed.

If only the horizontal two-dimensional degrees of freedom are considered, collisions can either dissipate or gain energy, depending on the amount of vertical energy grains have and the restitution coefficients. This idea was exploited in Ref. \cite{AlainBarrat}
in which the restitution coefficient was a random variable with possible outcomes larger than one. That model, however, lacked of an energy scale and the total energy of the system performs a random walk, not reaching a steady state. In the Q2D system, the vertical energy scale of the grains is fixed by the vibration parameters and so is the typical energy that is transferred from the vertical to the horizontal degrees of freedom. We propose a two-dimensional hard disk model, in which collisions are characterized by a constant restitution coefficient $\alpha$ and an extra velocity $\Delta$ that is is added to the relative motion. This extra velocity  points outwards in the normal direction $\hat{\sigma}$ as required by conservation of  angular momentum \cite{Lutsko2004}.
The collision rule for the post-collisional velocities reads
\begin{align}
 \Label{col.rule}
{\bf v}_1^*={\bf v}_1 -\frac{1}{2} (1+\alpha) ({\bf v}_{12}\cdot\hat\sigma) \hat \sigma -\hat\sigma \Delta ,\\
{\bf v}_2^*={\bf v}_2 +\frac{1}{2} (1+\alpha) ({\bf v}_{12}\cdot\hat\sigma) \hat \sigma  +\hat\sigma \Delta ,\nonumber
\end{align}
where ${\bf v}_{12}={\bf v}_1-{\bf v}_2$ is the relative velocity, $\hat\sigma$ points from particle 1 to 2, and particles are approaching if ${\bf v}_{12}\cdot\hat{\sigma}>0$.

With this set of collision rules, momentum is conserved, but energy is not conserved. 
The energy change in a given collision is
\begin{align}
E^*-E&= \frac{m}{2}
\left({\bf v}_1^{*2}+{\bf v}_2^{*2}- {\bf v}_1^2+{\bf v}_2^2  \right) \\
&=
m\Delta^2+m({\bf v}_{12}\cdot\hat\sigma)\alpha\Delta -
m({\bf v}_{12}\cdot\hat\sigma)^2 \frac{1-\alpha^2}{4} .
\end{align}
Considering a Maxwellian velocity distribution, absence of velocity correlations and static pair correlation function at contact $\chi$, the energy dissipation rate per unit area,  that should be included in the hydrodynamic equations,  is
\begin{equation}\Label{deltae}
G=-\frac{\omega(\rho,T)}{2}\left[m\Delta^2+\alpha\Delta \sqrt{\pi m T} -T(1-\alpha^2)\right] ,
\end{equation}
where $\omega(\rho,T)=2\rho \sigma \chi\sqrt{\pi T/m}$ is the collision frequency and the prefactor 1/2 compensates the double counting of collisions. 

The resulting expression of $G$ has the remarkable property that it is factorized into two terms that depend only on $\rho$ and $T$, respectively. This feature is a result of energy being injected and dissipated at collisions but not on the particular way of the  collision rule \reff{col.rule}. As a consequence, the stationary temperature, $T^{st}$, is density independent and is given by 
\begin{equation}\Label{eqn:Tst}
T^{st}= \frac{\pi \alpha^2}{4 (1 - \alpha^2)^2}\,\, \left(1 + \sqrt{1 + \frac{4 (1 - \alpha^2)}{\pi \alpha^2}}\right)^2\,\Delta^2.
\end{equation}
The stationary temperature diverges in the elastic limit ($\alpha\to1$) as energy is injected in every collision but no dissipation takes place. This divergency is correctly reproduced in simulations (see Sect. \ref{sec.md}). In order to obtain a finite temperature in the elastic limit, both the dissipation and the energy injection must vanish ($\alpha \to 1$, $\Delta \to 0$, and $\Delta/(1-\alpha)\to{\rm const.}$). 

In hard sphere models, the pressure is the temperature times  a monotonic function of the density. As in the proposed model (thermostated at collisions) the stationary temperature is density independent, the pressure increases monotonically with density and no effective negative compressibility can be produced~\cite{Cartes}. Therefore, effective two-dimensional collisional models are not able to reproduce the solid-liquid transition in Q2D systems. Our event driven simulations show that indeed the system in stable for all densities and inelasticities, giving rise to stationary homogeneous states.

The factorization of $G$ also implies that $G_\rho=0$ and therefore many expressions of the previous sections simplify. In particular the dissipative sound speed reduces to $c_d=\sqrt{p_\rho}$, the isothermal sound speed. Also, the hydrodynamic matrix is positive definitive  for any parameters and all modes are stable.

Finally, as advanced in Sect. \ref{sec.Skw.granular} the dissipation terms are proportional to the inelasticity. Indeed, when the partial derivative $G_T$  is computed and the stationary temperature is substituted, its dimensionless form is proportional to the inelasticity $1-\alpha$. 

\section{Numerical results of the collisional model} \Label{sec.md}

The effective 2D collisional model is simulated using the event driven algorithm for hard disks, considering the collision rule \reff{col.rule}. The disk diameter $\sigma$, particle mass $m$, and the energy injection parameter $\Delta$ are used to fix length, mass and time units. Simulations are done for systems of different restitution coefficients $\alpha$ and global density $\rho$. 
In this case we compute directly the intermediate scattering function, F({\bf k},t), that for $k\neq0$ reads 
\begin{equation}
F({\bf k},t)= \frac{1}{N}\left\langle\sum_{i,j } \exp[-i{\bf k}\cdot({\bf r}_i(t)-{\bf r}_j(0))]\right\rangle .
\label{Fkt}
\end{equation}
For isotropic steady states, as those reached for our model, $F({\bf k},t)=F(k,t)$, where $k=|{\bf k}|$. 
The reason to compute $F(k,t)$ instead of $C({\bf r},t)$ is that 
such  expression of the intermediate structure factor is well suited for numerical calculations, as it deals 
with analytical functions (exponentials) of the positions of the particles. Such exponentials have to be 
calculated  at regular instants of time $t$ for several $k$ vectors. Then, time correlations must be performed
to obtain $F({ k},t)$. On the contrary, the real space correlation functions, defined in Eq.~(\ref{CAB})
either require handling a delta function, or either performing averages over certain spatial domains in order to obtain coarse grained densities. 
Once $F({ k},t)$ is calculated, we carry out numerical Fourier transforms in time to compute $S({ k},\omega)$. 
To explore different wavevectors, different box sizes (and number of grains  $N$) were used to increase the wavelength and also, different wavevectors were analyzed simultaneously for a given system size. Finally, different aspect ratio $L_x/L_y$ were explored. 
In all cases, the system was verified to be stable and statistically homogeneous and,  for a given wavevector,  the computed physical quantities are independent of system size or aspect ratio.

The systems are initialized with a homogeneous distribution in space and velocities are sorted according to a Maxwellian distribution at the theoretical temperature $T^{st}$ \reff{eqn:Tst}. Then, the system is let to relax until a stationary state is reached.  Figure \ref{fig.Tst} shows the stationary temperature obtained in simulations compared with the predicted value. The agreement is excellent even for large inelasticities ($\alpha$ close to 0), where the hypothesis of Maxwellian distribution or absence of velocity correlations are expected to fail. Note that there is a small density dependence (few percents) on the stationary temperature implying that either there are velocity correlations or the velocity distribution depends on density.

\begin{figure}[htb]
\begin{center} \includegraphics[angle=270,width=.9\columnwidth]{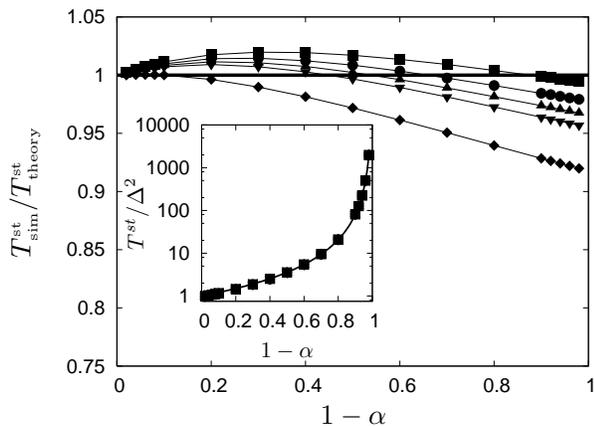} \end{center}
\caption{Stationary temperature divided by the theoretical value as a function of the inelasticity $1-\alpha$ for different densities. The temperatures are computed in molecular dynamic simulations of the collisional model described by the collision rule~\reff{col.rule}  while the theoretical values are obtained assuming  Maxwellian distributions and absence of velocity correlations. From top to bottom 
$\rho=$0.1, 0.2, 0.3, 0.4, and~$\rho=$0.8.  Inset: Theoretical (continuous line) and simulational (symbols) dimensionless stationary temperature $T^{st}/\Delta^2$ as a function of the restitution coefficient. The values for different densities collapse on the scale of the figure.}
\Label{fig.Tst}
\end{figure}

In the stationary state, the density fluctuations are obtained for different wavevectors, computing the intermediate scattering function $F(k,t)$ and the dynamic structure factor $S(k,\omega)$, that is shown in Fig. \ref{fig.Skw} for two different wavevectors. The two regimes (inelastic and quasielastic) are clearly seen: in the first case, only two peaks (sound modes, coming from Eq. (27)) are present while in the second case the three peaks (sound and heat modes) are visible.
In all cases, no more than  three peaks are observed showing that the pseudo-hydrodynamic model describes correctly the dynamics and it is not necessary to introduce additional kinetic modes \cite{9mom,BreyHydroValid}.

\begin{figure}[htb]
\begin{center} 
\includegraphics[angle=270,width=.9\columnwidth]{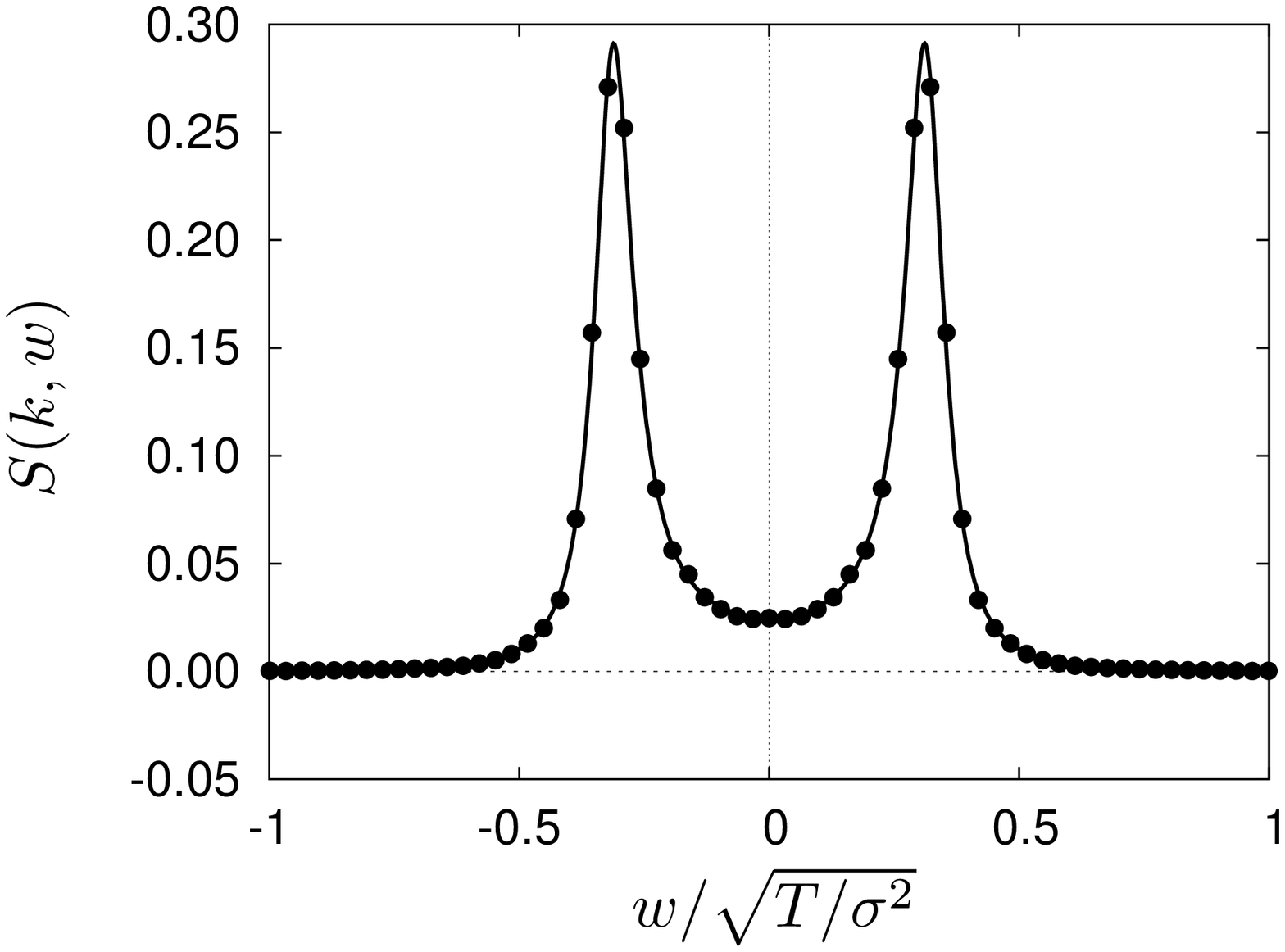}
\includegraphics[angle=270,width=.9\columnwidth]{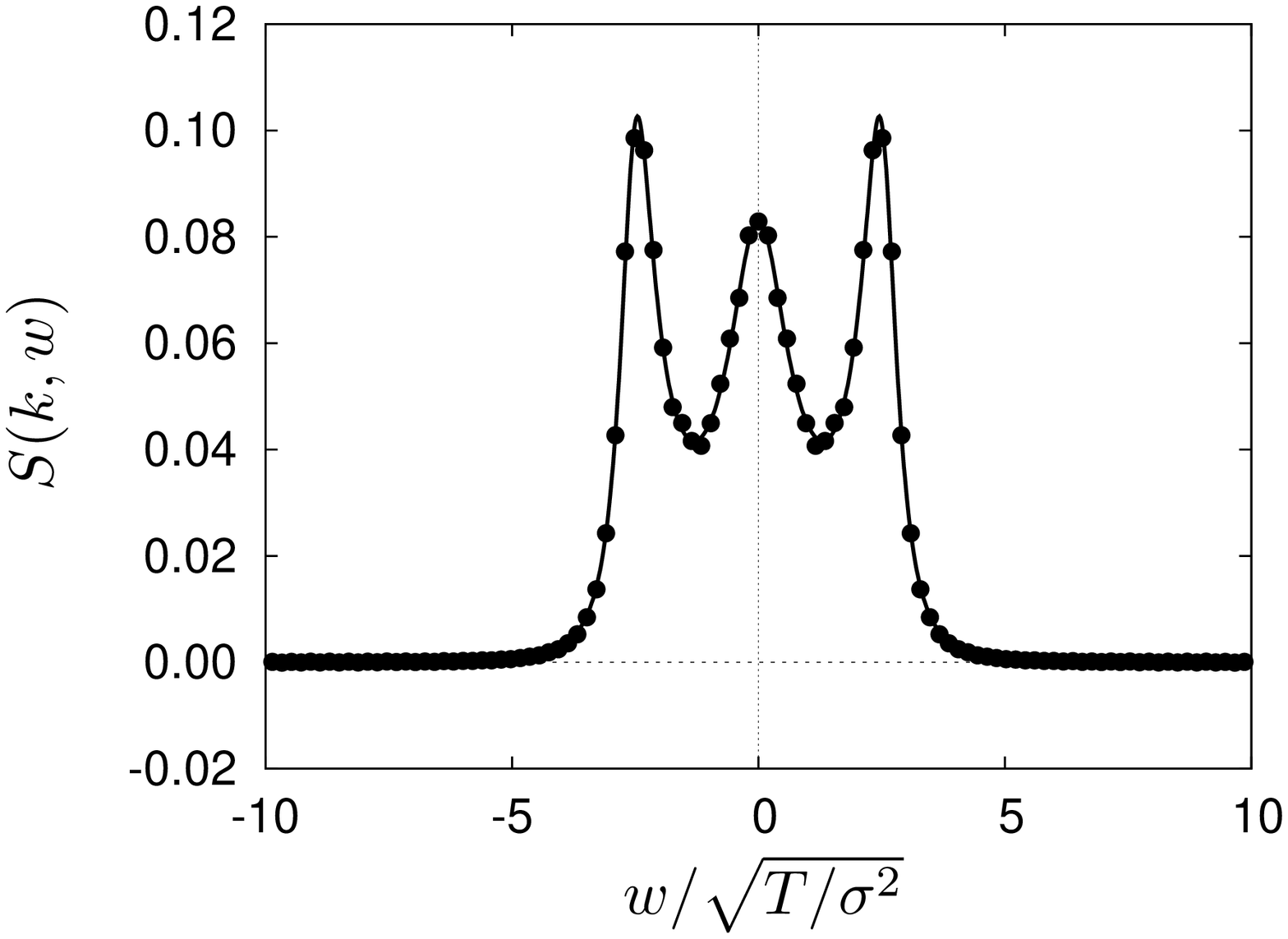}
\end{center}
\caption{Dynamic structure factor. The density and restitution coefficients are  $\rho=0.4$ and $\alpha=0.94$, respectively, while the wavevectors are $k\sigma =0.01$ (top) --corresponding to the dissipative regime-- and $k\sigma =0.06$ (bottom) --quasielastic regime.}
\Label{fig.Skw}
\end{figure}

In what follows we focus on the intermediate density $\rho=N\sigma^2/(L_x L_y)=0.4$; similar results are obtained for other densities.
To analyze the data, instead of fitting the position and width of the peaks in $S(k,\omega)$, we directly fit $F(k,t)$ with \reff{fit.Fkt} to obtain $\gamma$, $\omega_B$, $\widetilde{\Gamma}$, and $\widetilde{D}_T$ as a function of $k$ for different restitution coefficients.

\subsection{Transport coefficients and thermodynamic properties}
The pseudo-hydrodynamic equations must be supplemented by the equation of state $p(\rho,T)$, energy injection rate $G(\rho,T)$ and transport coefficients $\kappa$, $\eta$, and $\eta_V$. The computation of these require the analysis of the associated Enskog equation to find first the stationary distribution to compute $p$ and $G$. Second, the Enskog equation must be analyzed using the Chapman-Enskog method to compute the transport coefficients. The purpose of this article is to analyze the hydrodynamic models of granular matter rather than performing a kinetic theory description. There are numerous attempts to compute transport coefficients of granular fluids under different conditions, for example the homogeneous cooling state in two and three dimensions~\cite{IHS3D1,IHS3D2,IHS3D3}, the randomly driven gas~\cite{Twan}, the uniform shear flow~\cite{USFSantos,QHS}, and others. 
It has become evident that the transport coefficients and equations of state depend strongly on the energy injection mechanism, the reference state and not only on the restitution coefficient.  Therefore, previous predictions of transport coefficients or equations of state are not valid for the collisional model presented in this article. Even the first inelasticity correction is not valid.

Considering the above discussion, in order to make quantitative comparison with the simulation results, we will use quasielastic values. That is, the transport coefficients and the equation of state are those of the elastic fluid and only the energy injection rate is computed considering the inelasticity \reff{deltae}.  Expressions for these functions are given in  Appendix \ref{appendix.quasielastic}.

Using the numerical value of the pseudo-hydrodynamic matrix at each wavevector, we perform a full diagonalization of it. By this procedure we obtain $\omega_B$, $\widetilde{\Gamma}$, and $\widetilde{D}_T$ as a function of the wavevector, results that are compared with those obtained from the molecular dynamics simulations. 

\subsection{Sound modes}
In the whole range of inelasticities the sound modes are visible, being possible to obtain $\omega_B$ and $\widetilde{\Gamma}$ with good accuracy. As predicted, the sound frequencies go linearly with $k$ when the wavevectors are either in the inelastic or in the quasielastic regime, described in Sect. III and IV, respectively. Figure \ref{fig.GammaOmegaB} shows the sound velocity $\omega_B/k$ as a function of $k$ for a series of restitution coefficients. At small $k$, the sound velocity takes a constant value that tends to the dissipative speed $c_d$ while at large $k$ it takes a different value that approaches the adiabatic speed $c_s$, recovering the inelastic and quasielastic regimes discussed in the previous section. 

\begin{figure}[htb]
\begin{center} 
\includegraphics[angle=270,width=.9\columnwidth]{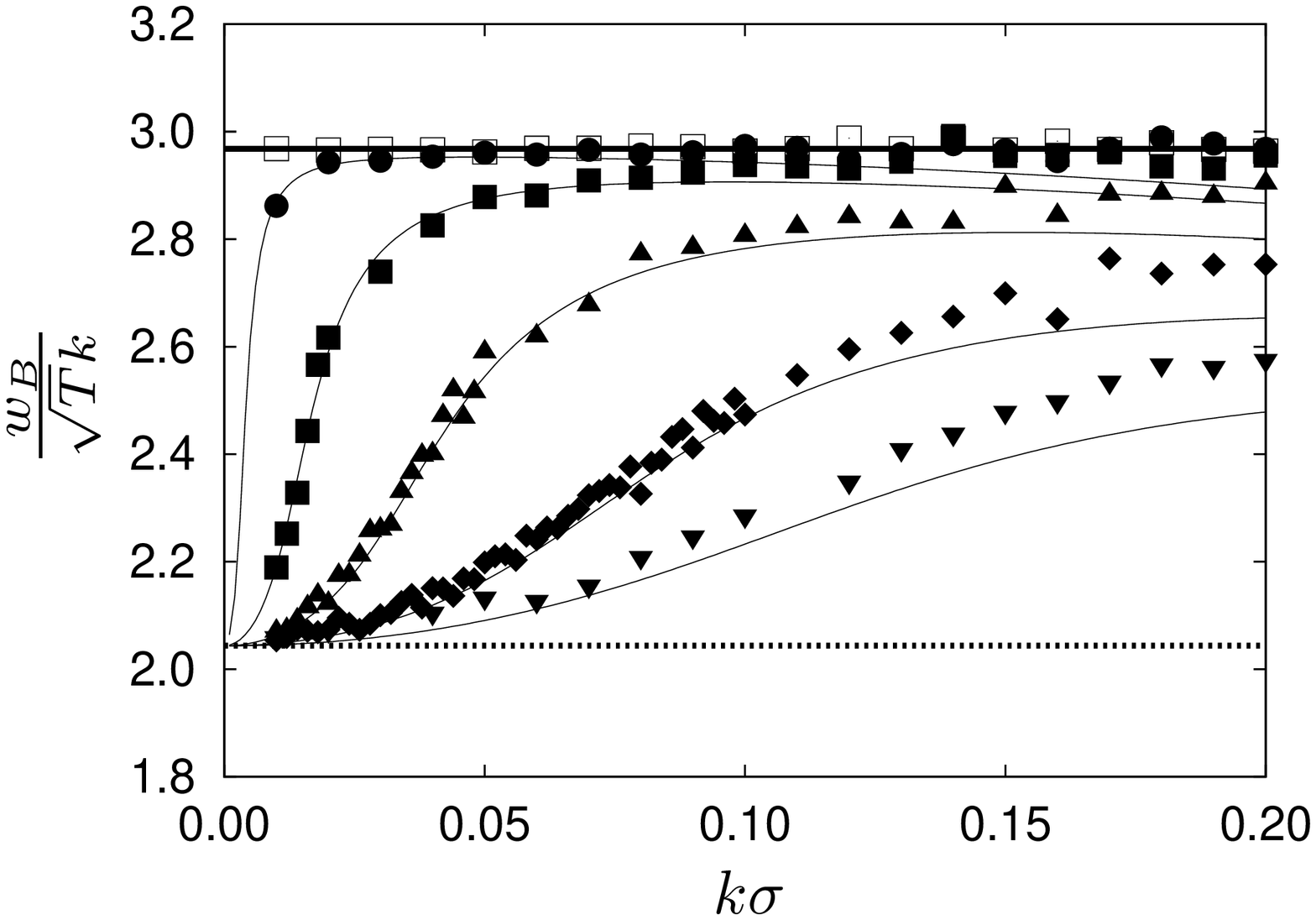}
\includegraphics[angle=270,width=.9\columnwidth]{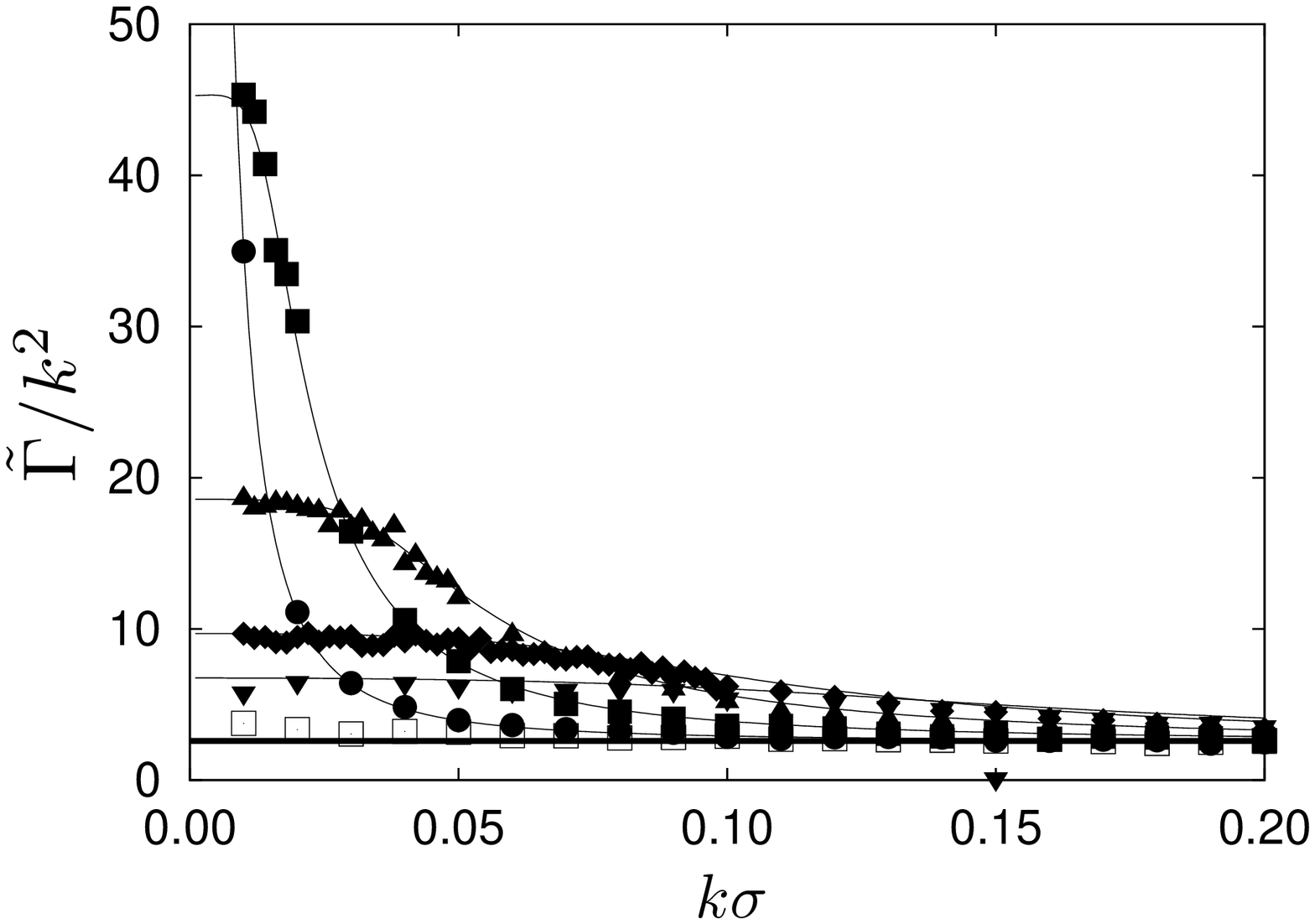}
\end{center}
\caption{Dimensionless sound velocity $\omega_B/k$ (top) and sound damping constant $\widetilde{\Gamma}/k^2$ (bottom) as a function of the dimensional wavenumber $k\sigma$ for different restitution coefficients
 ($\alpha$=1.00 $\Box$, 0.99 $\large \bullet$, 0.96 $\blacksquare$, 0.90$\blacktriangle$, 0.80 $\blacklozenge$ and  $0.70$ $\blacktriangledown$). Points are the results of the simulations and the lines the theoretical predictions using the quasielastic transport coefficients.
The solid horizontal lines are the predictions for an elastic fluid using Enskog theory: adiabatic velocity $c_s=2.968$ (top, solid line), dissipative (isothermal) velocity $c_d=2.044$ (top, dotted line), and sound damping constant $\Gamma=2.954$ (bottom).}
\Label{fig.GammaOmegaB}
\end{figure}

For small $k$, the width $\Gamma$ of the sound mode is proportional to $k^2$ as predicted. Therefore, Fig. \ref{fig.GammaOmegaB} shows $\widetilde{\Gamma}/k^2$ for a series of restitution coefficients. A similar crossover from the dissipative regime for small $k$ and the quasielastic regime for large $k$ is observed.  At large values of $k$ the sound damping constant tends to the quasielastic value which is almost independent of the restitution coefficient, while for small $k$, in the inelastic regime, the sound damping constant diverges as $1/(1-\alpha)$. Note that the elastic limit is singular as the crossover wavevector vanishes as well. Therefore, as expected, no divergence is obtained in the elastic case.

In Fig. \ref{fig.GammaOmegaB} we present as solid lines the predictions of the hydrodynamic equations. They are  obtained by full diagonalization of the hydrodynamic matrix $\mathsf{M}$ using the quasielastic transport coefficients described in Appendix \ref{appendix.quasielastic}. The agreement is excellent showing that both the hydrodynamic model is appropriate and that the quasielastic transport coefficients give a good approximation of the dynamics of the model, at least in the range of inelasticities presented in the figure.

The crossover wavevector $k_0$ between the inelastic and quasielastic regimes is obtained fitting the sound velocity to a Lorentzian 
\begin{equation}
\omega_B/k=c_s-\frac{c_s-c_d}{1+k^2/k_0^2} \label{fit.lorentzian}.
\end{equation}
Figure \ref{fig.k0} shows that $k_0$ is linear with the inelasticity $1-\alpha$ as predicted in Eq.~(\ref{k0.crossover}). A linear fit gives 
$k_0^{\omega_B} =  (0.530\pm0.005)(1-\alpha)/\sigma$. In the case of the sound damping, the crossover wavevector is obtained with a Lorentzian fit similar to (\ref{fit.lorentzian}) except that in this case all coefficients are free. The result, shown in Fig. \ref{fig.k0} is linear with $1-\alpha$ and a fit gives 
$k_0^{\Gamma} = (0.525 \pm 0.005) (1-\alpha)/\sigma$ which coincides with the value obtained using $\omega_B$. Both results should be compared with the prediction \reff{k0.pred} that, simplifies when $G_\rho=0$ to
\begin{equation}
k_0 = \left[\frac{c_V^2 \rho^4}{4c_V \rho^2 p\, p_\rho p_T -2 p^2 p_T^2-c_V^2 \rho^4 p_\rho^2}\right]^{1/4} G_T .\Label{k0.crossover}
\end{equation}
   Using the transport coefficients for quasielastic hard disks, we get a linear dependence with inelasticity 
$k_0=0.64 (1-\alpha)/\sigma$, that compares well with the simulations. Note that the value does not need to agree as the procedures to obtain $k_0$ are not exactly equivalent.

\begin{figure}[htb]
\begin{center} 
\includegraphics[angle=270,width=.9\columnwidth]{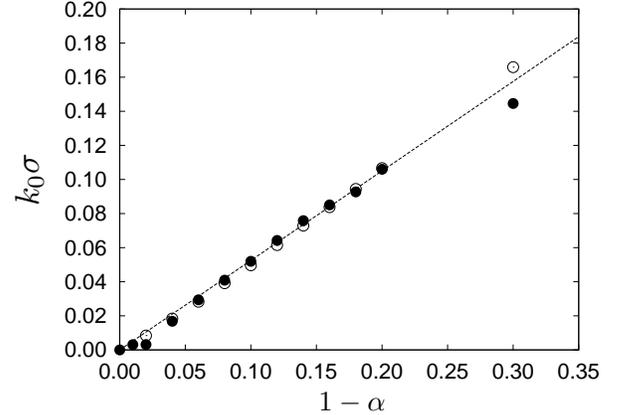}
\end{center}
\caption{Crossover wavevector for different inelasticities computed from Lorentzian fits of the sound speed ($\bigodot$) and sound damping constant ($\bullet$). The dotted line is the result of a linear fit, 
$k_0 = (0.525 \pm 0.005)(1-\alpha)/\sigma$. }
\Label{fig.k0}
\end{figure}

\subsection{Heat mode}
In the quasielastic regime the heat mode is well defined and the fitting procedure gives accurate values for $\gamma$ and $\widetilde{D}_T$.
In the inelastic regime (high inelasticities and  low wavevectors), 
the heat mode can be hidden by the sound modes as shown in 
Fig. \ref{fig.Skw} and the dynamics is represented by two modes only. 
However, using the numerical data from the simulations it is possible 
to force a fit $F(k,t)$  with three peaks using Eq.~(\ref{fit.Fkt}). 
Fig.~\ref{fig.gammamenos1} shows the relative amplitude of the heat peak 
compared to the sound peaks computed as the ratio between their areas 
$\gamma-1$. It is clear that in the inelastic regime the signal to noise 
ratio is poor and the precision in the fitted values for the heat mode is 
low.
\begin{figure}[htb]
\begin{center} 
\includegraphics[angle=270,width=.9\columnwidth]{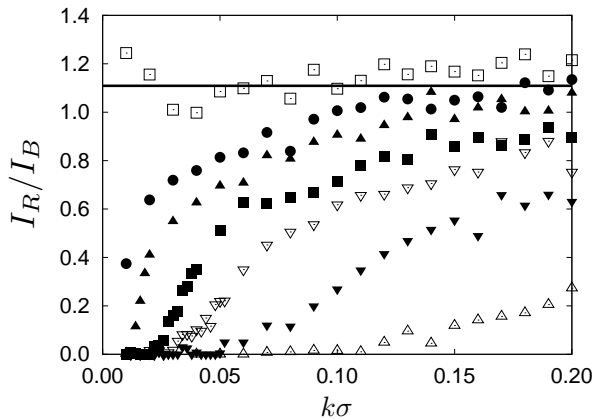}
\end{center}
\caption{Relative amplitude of the thermal peak to the sound peaks in the structure factor, $I_R/I_B=\gamma-1$, as a function of the dimensional wavenumber $k\sigma$ for different restitution coefficients 
($\alpha$=1.00 $\boxdot$, 0.98 $\bullet$, 0.96 $\blacktriangle$, 0.92 $\blacksquare$, 0.88 $\bigtriangledown$, 0.80 $\blacktriangledown$ and $\alpha =$0.60 $\triangle$). The solid horizontal line is the prediction for an elastic fluid using Enskog theory: $\gamma-1=1.109$. }
\Label{fig.gammamenos1}
\end{figure}

Figure \ref{fig.DT} presents the fitted values of 
the width of the thermal peak, $\widetilde{D}_T$. As expected, deep into the inelastic regime, it is not possible to obtain $\widetilde{D}_T$ with precision. In the quasielastic regime (large wavevectors) $\widetilde{D}_T$ shows a quadratic dependence with $k$. However, from this tendency it is not possible to obtain the  homogeneous dissipation $G_T$ by extrapolating it to $k\to0$. This is due to an increase of $\widetilde{D}_T$ when decreasing $k$ in the inelastic regime, as predicted in \reff{mode.thermal.inelastic} for which it can be verified that the coefficient of the $k^2$ term is negative (see Appendix~\ref{appendix.quasielastic}). The position of the minimum of  $\widetilde{D}_T$ is of the order of the crossover wavevector $k_0$.

Again, the comparison with the theoretical eigenvalues computed using the quasielastic transport coefficients is excellent for this range of inelasticities.

\begin{figure}[htb]
\begin{center}
\includegraphics[angle=270,width=.9\columnwidth]{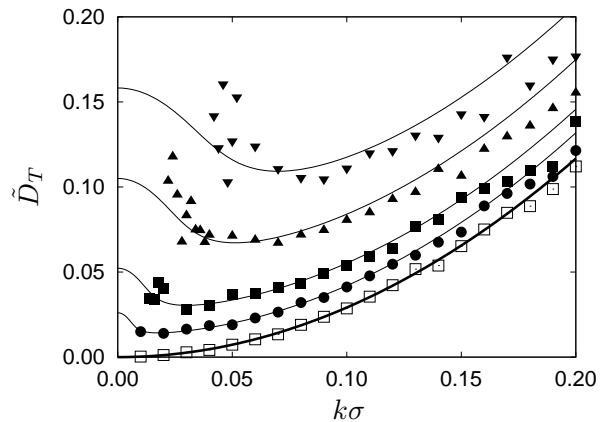}
\end{center}
\caption{Width of the termal peak $\widetilde{D}_T$ as a function of the dimensional wavenumber $k\sigma$ for different restitution coefficients ($\alpha$=1.00 $\boxdot$, 0.98 $\bullet$, $0.96$ $\blacksquare$, $0.92$ $\blacktriangle$ and $\alpha=0.88$ $\blacktriangledown$). Points are the results of the simulations and the lines the theoretical predictions using the quasielastic transport coefficients.}
\Label{fig.DT}
\end{figure}

\subsection{Transverse mode} \label{sect.transverse}
The transverse dynamics is much simpler as it decouples from the longitudinal modes. As shown in Eq.~(\ref{DeltaM}) the transverse eigenvalue is $\lambda_\perp=k^2\nu=k^2\eta/\rho$. The transverse eigenvalue is obtained from the simulations computing the self-correlation function of the transverse current 
\begin{equation}\Label{velocity.perp}
{\bf j_{\perp}}({\bf k},t)=  \sum_{i=1}^N(1-\widehat{{\bf k}}\widehat{{\bf k}})\cdot{\bf v}_i e^{-i {\bf k}\cdot{\bf r}_i(t)} ,
\end{equation}
where $\widehat{{\bf k}}={\bf k}/k$. The correlation function indeed decays exponentially, allowing the extraction of the transverse eigenvalue.
In the range $k\sigma\leq 0.2$ the transversal eigenvalues are quadratic with $k$ and the resulting viscosities are presented in Fig. \ref{fig.viscosity} for different inelasticities. The elastic value agrees with the Enskog prediction in 2D and it takes smaller values as the inelasticity is increased. A linear fit gives
\begin{equation}
\nu = (1.314\pm0.004)-(0.37\pm0.01) (1-\alpha)
\end{equation}
in the range $0.1 \leq \alpha \leq 1$, that is almost up to the plastic limit.
 
\begin{figure}[htb]
\begin{center}
\includegraphics[angle=270,width=.9\columnwidth]{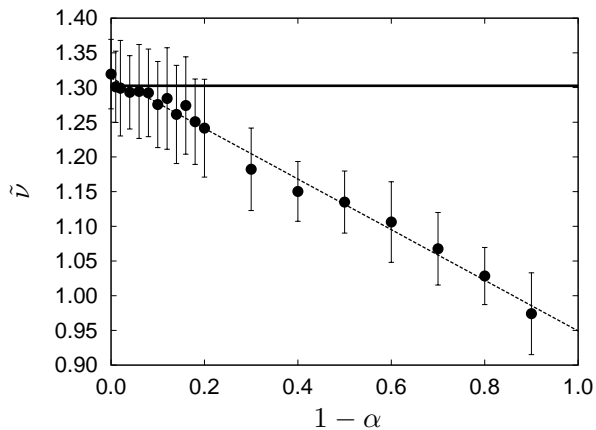}
\end{center}
\caption{Dimensionless kinematic viscosity $\tilde{\nu}=\eta/(\rho/\sqrt{T})$as a function of the inelasticity $1-\alpha$ obtained from simulations of the collisional model (symbols). The dashed line is a linear fit  
$\tilde{\nu} = (1.314\pm0.004)-(0.37\pm0.01) (1-\alpha)$
and the solid line is the theoretical value for an elastic fluid obtained form the Enskog value $\tilde{\nu}_{\rm elastic}=1.303$. }
\Label{fig.viscosity}
\end{figure}

\section{Conclusions} \Label{sec.conclusions}

We study the dynamics of a granular medium subjected to a bulk energy injection. Making simple generic assumptions on the injection method (momentum conservation and Galilean invariance) the pseudo-hydrodynamic equations are written. These describe the dynamics of the conserved density and velocity fields and the non-conserved temperature field. 

The fluctuations about the stationary state are analyzed  and described in terms of the dynamic structure factor and the corresponding eigenvalues of the inelastic hydrodynamic matrix. The dynamics near the stationary state is characterized in term of the following modes: the viscous mode that decouples as usual from the rest, the sound modes and the heat mode. Two regimes are clearly distinguished. First the dissipative regime in which the heat mode is suppressed and the effective dynamics is reduced to only two fields and, second, the quasielastic regime in which the heat mode is visible. The crossover wavevector is proportional to the dissipation. In the dissipative regime the sound speed is isothermal and the sound damping constant becomes large, diverging in the limit of small dissipation and small wavevectors; the elastic limit is singular as it can be obtained by only making first the dissipation small and only later the wavevector can be small. In the quasielastic regime the sound velocity is the adiabatic one and the sound damping constant is similar to the elastic value. 

The general predictions are compared with a collisional model that mimics the effective two-dimensional dynamics of a horizontally shallow three-dimensional system. In the shallow system the box is  vertically vibrated and the energy is transferred from the box to the grains and later to the two-dimensional degrees of freedom through grain-grain collisions. To model this, we consider a purely two-dimensional granular fluid and, in collisions, a fixed additional separation velocity is added to the postcollisional velocities of each grain. A stationary temperature is reached that depends on the restitution coefficient and this added velocity.

Molecular dynamics simulation of this model confirm the qualitative description of the hydrodynamic modes for the fluctuations. Besides, using the transport coefficients of the elastic fluid plus the energy injection function computed for the model, there is a very good numerical agreement with the theoretical predictions. It is difficult, however, to obtain the transport coefficients from a fit of the simulational eigenvalues. This is due to numerical accuracy and the fact that there are too many unknowns to be fit: the transport coefficients, the equation of state and the energy injection rate. In the case of the transverse mode it is possible to fit the viscosity obtaining its dependence with the inelasticity. 

\acknowledgments{The authors thank Ana Asenjo for useful comments. 
The research is partially supported by Spanish grants MODELICO and ENFASIS, the  Fondecyt grants 1100100 and 1120775, 
and {\em Proyecto Anillo  ACT 127}. }

\appendix

\section{Dynamic Structure Factors} \label{app.Skw}
In a system composed of $N$ grains  in a  volume V (global density $\rho=N/V$), the local density field is defined as
\begin{equation}\Label{density}
\rho({\bf r},t) = \sum_{i=1}^N\delta ({\bf r}-{\bf r}_i(t)),
\end{equation}
where ${\bf r}_i(t)$ is the position of the $i$-th particle at time $t$. The densiy-density  correlation function is
\begin{equation}\Label{densitymod}
C_{\rho\rho}({\bf r},t)=\langle \rho({\bf r}+{\bf r}',t+t') \rho({\bf r}',t')\rangle - \rho^2.
\end{equation}
For the velocity correlation function, which is a tensorial quantity, the dynamic variable is
\begin{equation}\Label{velocity}
{\bf j}({\bf r},t)= \sum_{i=1}^N{\bf v}_i \delta ({\bf r}-{\bf r}_i(t)).
\end{equation}

By taking the Fourier transform in space of $C_ {AB}({\bf r},t)$ 
we obtain the so called intermediate scattering function. For the density-density case, it reads
\begin{align}
F({\bf k},t)&= \rho^{-1} \int \,dr\, e^{-i{\bf k}\cdot{\bf r}} C_{\rho\rho}({\bf r},t)\\
&= \frac{1}{N}\left\langle\sum_{i,j } \exp[-i{\bf k}\cdot({\bf r}_i(t)-{\bf r}_j(0))]\right\rangle -
 (2\pi)^3 \rho \delta({\bf k}). 
\label{Fkt2}
\end{align} 
The last term containing a Dirac delta at ${\bf k}=0$ only represents the mass conservation, that will be dropped from here on.

Subsequently, one can perform a Fourier transform in time variable of  
$F(k,t)$ to obtain the dynamic structure factor, $S(k,\omega)$. Such structure factor is a fundamental tool
in the study of fluid systems, like gases, liquids, polymers, 
and colloids~\cite{Berne}. 
The reason is that, at long wavelengths, $k\to 0$,  $S(k,w)$ encodes many equilibrium and non equilibrium properties of the fluid.
At long wavelengths, it allows full evaluation, by the so called Landau-Plazceck approximation, where the evolution of the fields is given by the hydrodynamic equations.
In this regime, $S(k,\omega)$ shows three Lorentzian peaks. Its expression is \cite{Boon} (we quote it here for  reference and comparison 
with the inelastic case)
\begin{align}\Label{skw}
S ({ k},\omega)/S(k) =&    \frac{\gamma-1}{\gamma} \frac{2 D_T k^2}{\omega^2+(D_T k^2)^2} \nonumber \\
 &  + \frac{1}{\gamma}\frac{\Gamma k^2}{(\omega \pm c_s k)^2 + ( \Gamma k^2)^2} .
\end{align}
 The first term represents a peak located at $\omega=0$ (called Rayleigh peak) and appears as a consequence of energy conservation.
It has  a width given by $D_T= \kappa/(\rho c_V)k^2$,
where $\kappa$ is the heat conductivity of the fluid, $\rho$ is the average density and $c_V$ is the 
specific heat at constant volume. Such a peak carries the information about the entropy evolution in the system.

There are two other peaks, Brillouin peaks (represented by the symbol $\pm$ in the denominator), symmetric respect to that of $\omega=0$, located at $\omega=\pm i c_s k$, where $c_s$ is the adiabatic speed of sound ($c_s=\sqrt{(c_p/c_V)(\partial p/\partial \rho)}$, where $c_p$ is the specific heat at constant pressure) . 
Their presence is a consequence of the conservation of momentum and the inertia of  the system. 
The width of these peaks is $\Gamma k^2$, where $\Gamma$ is the sound damping constant, a combination of shear and bulk viscosities and the heat conductivity. 
Moreover, the area enclosed by the thermal peak divided by the area under the sound  peaks is $\gamma-1$, being $\gamma$ the adiabatic constant, so it yields the ratio of specific heats $c_p/c_v$.  

The expression at Eq.~(\ref{skw}) is obtained at the lowest order in the wave vector $k$. The next order 
term in $k$ gives the so-called {\em asymmetric} contribution to the Brillouin peaks that vanishes in the hydrodynamic limit, but have a finite contribution at finite $k$ \cite{Boon}. Such asymmetric peaks are considered in the paper. 
In summary, the measure of $S(k,\omega)$ in molecular fluids allows us to obtain transport coefficients and some thermodynamic properties.

\section{Thermodynamic properties and transport coefficients of the elastic hard disk fluid} \label{appendix.quasielastic}
Here we provide the expressions for the thermodynamic properties and transport coefficients of the elastic hard disk fluid (2D), used in the  comparison with the simulation results. Units are such that the grain diameter $\sigma$ and mass $m$ are set to one.

The thermodynamic properties of an elastic hard disk (2D) fluid can be obtained using the expressions of the equation of state \cite{EqState1,EqState2} and the specific heat at constant volume
\begin{align}
p & = \rho T \frac{\left(1+\frac{\rho^2\pi^2}{128}\right)}{\left( 1-\frac{\rho\pi}{4}\right)^2} ,\\
c_V & = 1 ,
\end{align}
where temperature is measured in energy units. From these it is straightforward  to obtain the adiabatic and dissipative sound speeds, and the adiabatic constant
\begin{align}
c_s&=\sqrt{p_\rho+p_T^2/c_V\rho^2}  \approx 2.97 \sqrt{T} ,\\
c_d &=\sqrt{p_\rho} \approx 2.04 \sqrt{T} ,\\
\gamma-1 & = c_P/c_V \approx 1.11,
\end{align}
where, in the right equalities the reference density $\rho=0.4$ used in the simulations along the paper has been replaced.

The transport coefficients of the hard disk gas have been obtained using the Enskog transport equation \cite{GassTrCoef}, and are valid up to moderate densities
\begin{align}
\eta &= \frac{1.022}{\chi}\sqrt{\frac{T}{2\pi}} \left[1+(2\rho\chi) + 0.8729 (2\rho\chi)^2 \right]  \approx 0.52\sqrt{T} \\
\eta_V & = \frac{1.022}{\chi}\sqrt{\frac{T}{2\pi}} \left[ 1.246 (2\rho\chi)^2\right] \approx 0.26\sqrt{T}\\
\kappa & =  \frac{1.029}{\chi}\sqrt{\frac{2T}{\pi}} \left[1+\frac{3}{2}(2\rho\chi) + 0.8718 (2\rho\chi)^2 \right] \approx 2.46\sqrt{T},
\end{align}
where $\chi$ is the pair correlation function at contact given by
\begin{equation}
\chi = \frac{1-\frac{7}{16}\frac{\pi\rho}{4}}{\left(1-\frac{\pi\rho}{4}\right)^2} \approx 1.83,
\end{equation}
and, as before,  the reference density $\rho=0.4$ has been replaced in the right equalities.

Finally, some expressions that are used in the computations of the eigenvalues are
\begin{align}
D_T&=\frac{\kappa p_\rho\rho}{p_T^2+c_Vp_\rho\rho^2} \approx 2.92 ,\\
\Gamma &= \frac{\nu_l}{2}+\frac{\kappa p_T^2}{2c_V\rho(p_T^2+p_\rho \rho^2 c_V)} \approx 2.95 .
\end{align}

\end{document}